# Hierarchical stripe phases in IrTe$_2$ driven by competition between Ir dimerization and Te bonding

Jixia Dai[1,2], Kristjan Haule[1], J.J. Yang[3], Y.S. Oh[1,2], S-W. Cheong[1,2,3] and Weida Wu[1,2]

[1] Department of Physics and Astronomy, Rutgers University, Piscataway, NJ, 08854, USA

[2] Rutgers Center for Emergent Materials, Rutgers University, Piscataway, NJ, 08854, USA

[3] Laboratory for Pohang Emergent Materials and Department of Physics,
Pohang University of Science and Technology, Pohang 790-784, Korea

**Abstract**

Layered 5$d$ transition metal dichalcogenide (TMD) IrTe$_2$ is distinguished from the traditional TMDs (such as NbSe$_2$) by the existence of multiple CDW-like stripe phases and superconductivity at low temperatures. Despite of intensive studies, there is still no consensus on the physical origin of the stripe phases or even the ground state modulation for this 5$d$ material. Here, we present atomic-scale evidence from scanning tunneling microscopy and spectroscopy (STM/STS), that the ground state of IrTe$_2$ is a $q$=1/6 stripe phase, identical to that of the Se-doped compound. Furthermore, our data suggest that the multiple transitions and stripe phases are driven by the intralayer Ir-Ir dimerization that competes against the interlayer Te-Te bonding. The competition results in a unified phase diagram with a series of hierarchical modulated stripe phases, strikingly similar to the renowned "devil's staircase" phenomena.

PACS: 71.45.Lr, 71.27.+a, 74.55.+v

† To whom correspondence should be addressed: wdwu@physics.rutgers.edu

One major challenge in condensed matter physics is to understand the intriguing interplay between unconventional superconductivity and other modulated electronic ground states, e.g. anti-ferromagnetism, charge/orbital ordering, stripes, structure distortion, spin/charge-density wave (SDW/CDW), etc. Clarifying the microscopic nature of these phases is crucial for understanding the mechanism of unconventional superconductivity[1,2]. Layered TMD have been an active playground for exploring the interplay between CDW and superconductivity in the last several decades[3-6]. Yet it is still under debate what is responsible for the CDW transitions in these materials[7-10]. This exploration is further fueled by the recently discovered superconductivity[11-14] emerging from the suppression of a CDW-like stripe phase transition ($T_C \approx 280$ K) with $q=1/5$ along the (1,0,-1) direction in IrTe$_2$ (ref. 11, 12, 15). The qualitative similar phase diagram between IrTe$_2$ and other unconventional superconductors suggests that the driving mechanism of the stripe phase may hold the key to understand the superconductivity. Recent STM and transport studies suggest that the CDW-like transition in IrTe$_2$ is followed by another one near $T_S \approx 180$ K (when cooled) with further increase in electric resistivity[16], a phenomenon similar to the multi-step CDW transitions in the quasi-1D system NbSe$_3$ (ref. 17) and quasi-2D TaS$_2$ (ref. 6). Unlike these two systems, IrTe$_2$ is unique in terms of single-$q$ (i.e. stripe) modulations with unusually large hysteretic transitions[16], implying a distinctive but yet to be unveiled driving mechanism behind them.

Although many candidate mechanisms, ranging from Fermi surface nesting[11], Ir bonding instability[12,18], crystal field effect or interlayer bonding of Te $p$-orbitals[13,14,19], have been proposed based on various experimental observations, none of them explains all the experimental results satisfactorily. For example, the absence of CDW gap in angle-resolved photoemission spectroscopy (ARPES) and optical spectroscopy measurements excludes the Fermi surface instability mechanism[13,20]. The central question of the heated debate is whether Ir or Te plays the key role in the driving mechanism. This is partially due to the complication that the single-$q$ modulation of the stripe phases results in multi-domain states below the transition, which washes out the intrinsic properties of spatially-averaged experimental observation. Indeed, the atomic position of the $q=1/5$ stripe phase was unambiguously resolved with single crystal x-ray refinement on single domain sample[21]. Yet, this technique cannot address the emergence of short-range multiple modulations below $T_S$. Therefore, real-space resolving techniques, especially at low temperature, are necessary for addressing the fundamental mechanism.

Previous STM studies revealed that the second transition at $T_S$ is a hysteretic transition from a periodic soliton lattice ($q=1/5$) to a partially melted soliton state consisted of emergent stripe modulations of $q_n=(3n+2)^{-1}$, with gradual reduction of soliton density down to T = 50 K [16]. A natural extrapolation ($n\rightarrow\infty$) of this behavior indicates the ground state of IrTe$_2$ is a soliton-free $q=1/6$ modulated phase, in contrast to previous studies[11, 19, 22]. Thus, a clarification of the IrTe$_2$ ground state is imperative for understanding the fundamental mechanism that drives the transition. In this work, we performed low-temperature STM/STS studies of undoped, selenium- and platinum-doped IrTe$_2$. Our high resolution topographic and spectroscopic data provide strong evidence that the low temperature phase of IrTe$_2$ is a $q=1/6$ phase, the same as that in the Se-doped IrTe$_2$ (ref. 19). Combining the STM/STS measurements and first principle calculations, our results support that the competition between Ir-Ir dimerization and interlayer Te-Te bonding is the fundamental driving force of the hierarchical CDW-like phase transitions in IrTe$_2$. Such intriguing phenomena are closely related to the emergence of hierarchical modulations due to competing microscopic interactions in the famous "devil's staircase" phenomena[23].

In this study, we used single crystals of IrTe$_2$, IrTe$_{2-x}$Se$_x$ ($x=0.34$) and Pt$_x$Ir$_{1-x}$Te$_2$ ($x=0.04$ and 0.1) grown using Te flux, as described in ref. 19. For STM/STS measurements, an Omicron LT-STM with base pressure of $1\times10^{-11}$ mbar was used. Electrochemically etched tungsten tips were calibrated on clean Au (111) surface before STM experiments on IrTe$_2$ (Supplementary Fig. **S5**). The single crystal samples were mounted on the sample plates using conducting silver epoxy (H20E from EPO-TEK). After being introduced into the STM chamber, the samples were cleaved *in situ* and loaded into the STM scan-head within one minute. The temperature of the scan-head was held at 4.5 K by liquid helium. The differential conductance measurements were performed with the standard lock-in technique with amplifier gain $R_{gain}$ = 3 GΩ, modulation frequency $f$ = 455 Hz and modulation amplitude $V_{mod}$ = 5~10 mV. To model the LDOS measured by STS, we have carried out the density functional theory calculation, using the WIEN2K package[24]. The $p_z$ partial density of states was obtained by the DMFT-WIEN2K package[25], where spin rotations (in the presence of large spin-orbit coupling) are properly implemented.

The local coordinate system on Te atoms is chosen such that $p_z$ orbital points perpendicular to the surface Te-layer, which maximizes its coupling to the tunneling tip. We used the crystal structure for IrTe$_2$ in the 1/6 phase determined in ref. 26.

Figure **1a** and **1b** show the atomically resolved STM images of IrTe$_2$ and IrTe$_{2-x}$Se$_x$ ($x$=0.34), respectively. The contribution from electronic (charge) modulations was removed to reveal atomic positions of Te atoms (see Supplementary Fig. **S1**). At 4.5 K, the surface of IrTe$_2$ is dominated by uniform 1/6 modulation (Fig. **S2**) except for a small fraction of areas containing dilute solitons (Fig. **3a**). The dimer stripes pattern is clearly visible. Similarly, long range 1/6 modulation was observed (Fig. **1b**) on the surface of IrTe$_{2-x}$Se$_x$, even in the presence of Te/Se atomic disorder. The atomic scale inhomogeneity is removed by unit-cell averaging, resulting in images (Fig. **1c** & **1e**) showing practically identical dimer stripe pattern. This strongly suggests that the 1/6 modulation in these two compounds are the same. Furthermore, the "bond" length of Te dimers (the red dashed curves) is about 3.2 Å, which is ~18% shorter than that in the high temperature phase. This is in excellent agreement with the results of single crystal X-ray refinement of the Te dimers in both 1/5 and 1/6 phases[21, 26], further indicating the dimers in the 1/5 and the 1/6 states are of the same origin. (See Supplementary Fig. **S3** for STM images of the 1/5 state.) As shown in the simplified atomic structure model of 1/6 super-cell (Fig. **1d**), the Te atoms shift collectively to generate the 3+3' periodic structure. The second half of super-cell (3') is related to the first half (3) through symmetry operations of a mirror plus a translation along the horizontal plane. Therefore, the primary 1/6 peak is absent due to a structural factor cancellation, while the satellite 1/6 peak is observed near (010) as shown in the Fourier map (Fig. **1f**) and line profiles (Fig. **1g**), consistent with previous TEM studies of IrTe$_{2-x}$Se$_x$ (ref. 19). Accordingly, the primary electronic (charge) modulation is 1/3 (Supplementary Fig. **S1**).

The excellent agreement between the two 1/6 states in the undoped and the Se-doped IrTe$_2$ is further corroborated by STS measurements (i.e. differential tunneling conductance *dI/dV*), which probe local density of states (LDOS)[27]. Figure **2a** (**2c**) show the unprocessed STM topography of IrTe$_2$ (IrTe$_{2-x}$Se$_x$). The spatially averaged STS data taken on Fig. **2a** (**2c**) is shown in Fig. **2b** (**2d**). Over a wide energy range (+/-2 eV), the *dI/dV* spectra from these two compounds are qualitatively the same: sharing both a cliff-like particle-hole asymmetry near

Fermi level and a pronounced peak near 1.5 eV. Even on the areas of dilute solitons in IrTe$_2$, similar spectrum is observed between the solitons (red box in Fig. **2c** and the corresponding *dI/dV* in Fig **2g**), confirming previous conjecture that 1/6 phase is the soliton free limit of $q_n = (3n+2)^{-1}$ modulations[16]. High spatial resolution STS measurements reveal further variation of LDOS at different atomic rows. As shown in Fig. **2e**, the STS data of individual atomic columns (defined in Fig. **2f**) show LDOS variation, indicating that the LDOS of column 2 (which is above Ir dimers) is lower than those of column 1 and 3 between Fermi energy ($E_F$) and ~1.6 eV. These results are in qualitative agreement with the calculated LDOS of Te $p_z$ orbitals shown in Fig. **2g**, because they dominate the tunneling process due to orbital orientation. The difference between theory and experimental data may originate from energy dependence of tunneling matrix element.

However, the LDOS of soliton crest (averaged over the blue box in Fig. **3a**) is markedly different: showing a pronounced "hump" at the Fermi level accompanied by spectral-weight suppression from 0.5 eV and above (Fig. **3b**). The observed reduction of density of states near $E_F$ in the dimer phase agrees with previous first principle calculation[21]. Consistently, resistivity jumps were observed at the two transitions during cooling as the dimer density increases[16]. Additionally, there is no gap feature around zero bias in any of the above spectra, in agreement with previous angle-resolved photoemission spectroscopy (ARPES) and optical measurements[13, 20].

The striking similarity between the low temperature 1/6 stripe phases between undoped and Se-doped IrTe$_2$ clearly demonstrate that they share the same origin. Furthermore, the transition temperature ($T_C$) is enhanced smoothly with Se doping[19]. Together these observations suggest that Te plays a secondary, likely a competing role in the CDW-like transition. Indeed, a completely different scenario emerges in Pt-doped compound Pt$_x$Ir$_{1-x}$Te$_2$. At low doping as $x = 0.04$, the $q=1/5$ stripe phase is stabilized at low temperature (Fig. **3c**), indicating that the soliton melting transition ($T_S$) is suppressed. When the Pt concentration is further increased ($x=0.1$), transport measurements show that the CDW-like transitions is suppressed while superconductivity emerges in the vicinity[12, 13, 28]. Consequently, the high temperature trigonal symmetry (1×1 phase) is observed in our STM measurements of $x=0.1$ samples (Fig. **3e**), demonstrating the absence of the long range order of any stripe modulations.

Furthermore, the *dI/dV* spectrum measured on this 1×1 phase (Fig. **3f**) is very similar to that of soliton crests (undimerized stripes), confirming previous theoretical prediction that these undimerized stripes are essentially the high temperature phase[21]. The strong suppression of stripe phases by Pt doping is in sharp contrast to the enhancement of $T_C$ with Se doping, which provides compelling evidence that Ir dimerization plays the dominant role on driving the transition.

Compiled with previous studies, our STM results can be summarized in a unified phase diagram (Fig. **4**) showing the antagonistic behaviors of $T_C$ with Se- and Pt-doping. The salient feature of this phase diagram is the common ground state (1/6 phase) of Se-doped and undoped $IrTe_2$. Previous studies emphasize the importance of Te orbitals, especially interlayer Te-Te bonding (polymerization) in the high temperature 1×1 phase [13, 14, 19, 29]. However, it wasn't elucidated what breaks the Te-Te bonding below $T_C$. Using single-crystal x-ray refinement, Pascut *et al*. observed striking intralayer Ir-Ir dimer formation with >20% bond length reduction, stronger than that of Te-Te dimer (10-17%) in the 1/5 stripe phase. This bond length change of Ir-Ir dimer is even larger than the that in $CuIr_2S_4$ (ref. 30) and $VO_2$ (ref. 31-33) where dimerization is associated with their metal-insulator transitions. Clearly, it is the Ir dimerization that competes against the interlayer Te-Te bonding and drives the system into the dimer stripe phases. Competing microscopic interactions often leads to phenomena of emergent periodicities such as incommensurate density waves and devil's staircase (ref. 23, 34, 35). Therefore this mechanism naturally explains the appearance of $q$=1/5 phase and the subsequence $q_n$=$(3n+2)^{-1}$ stripe phases below 180 K, namely they are the staircase phases precede the ultimate ground state ($q$=1/6). Se-doping weakens the interlayer Te-Te bonding, resulting in an increase of $T_C$ and likely a suppression of temperature window of the $q_n$=$(3n+2)^{-1}$ stripe phases. In contrast, Pt substitution of Ir directly perturbs the dimerization, resulting in a dramatic suppression of $T_C$ (ref. 12, 13) and enhanced stability of the intermediate (1/5) phase. Interestingly, superconductivity appears near the suppression of long range ordering of stripe phase, indicating fluctuating Ir dimers might facilitate cooper pairing of electrons.

In summary, our results of atomic scale imaging and spectroscopic measurements suggest that the ground state for $IrTe_2$ is the $q$=1/6 state with the alternating dimer pattern, while doping into the Ir sites weakens the tendency for the system to form dimers. Nonetheless, comprehensive understanding of the ground states and their relation to the superconductivity would require further experimental and theoretical efforts. Due to the stripe and thus the domain formation, spatially averaged measurements, such as transport, quantum oscillation, ARPES and scattering experiments, would have extra complication with data interpretation. Measurements within a single domain would provide more straightforward information, such as the fermi surface, conductivity anisotropy and the unusual cross-layer quasi-2D electronic behavior. We expect the single domain phase to be achieved by straining the crystal while cooling through the transitions and/or repetitive thermal cycling, similar to the Fe-base superconductors [36]. Furthermore, by low Se doping, the transition temperature could be tuned around room temperature[19], which is highly desirable for potential applications.

We thank G. L. Pascut and V. Kiryukhin for sharing X-ray results and discussions. The work at Rutgers was supported by NSF Grants No. DMREF-1233349 and No. DMR-0844807. The work at POSTECH was supported by the Max Planck POSTECH/KOREA Research Initiative Program (No. 2011-0031558) through the NRF of Korea funded by the Ministry of Education, Science and Technology.

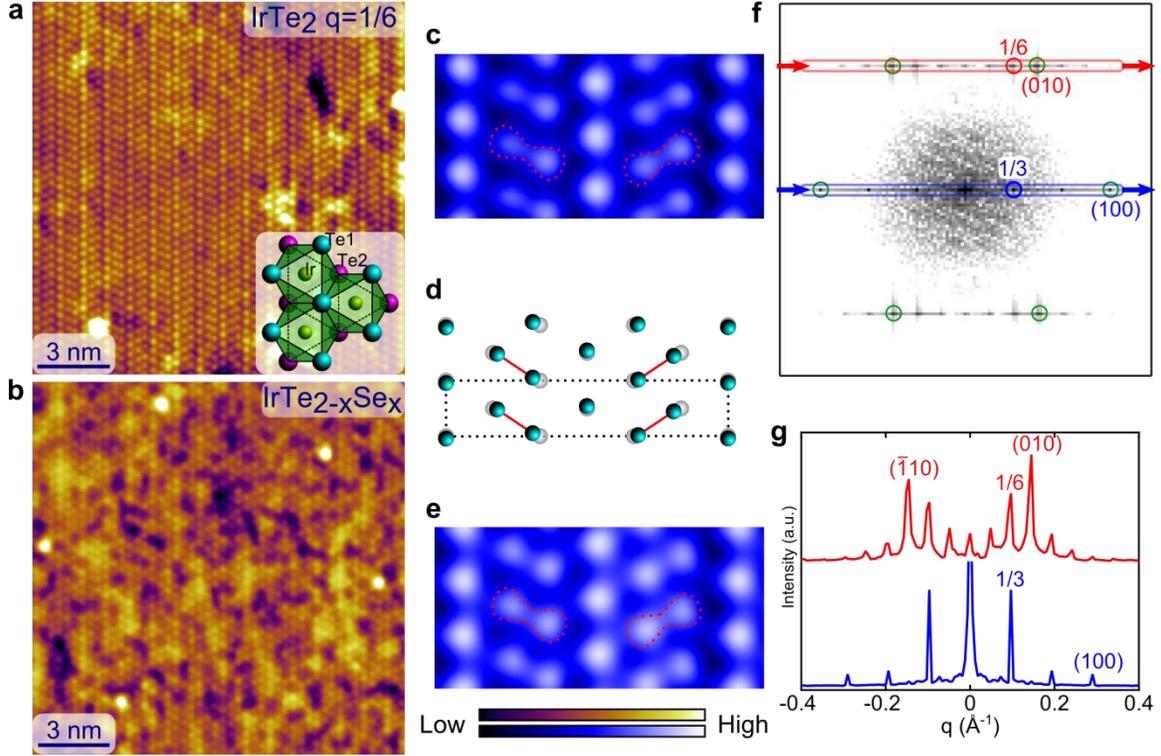

**FIG. 1** (Color online) Visualizing the Te-Te dimer stripes on the surface of IrTe$_2$ and IrTe$_{2-x}$Se$_x$. (**a**) and (**b**) topographic images of the $q=1/6$ states in IrTe$_2$ and IrTe$_{2-x}$Se$_x$ at 4.5 K without the 1/3 charge density modulation. The inset of (**a**) shows the crystal structure of IrTe$_2$. The atoms in the STM images correspond to the top Te layer (Te1). (**c**) and (**e**) are the unit-cell-averaged images of (**a**) and (**b**), respectively. Here the alternating Te-Te dimers are clearly visible (indicated by the red dashed curves). (**d**) Schematic diagram of the unit cell for the 1×6 structure, illustrating individual atomic shifts and the Te-Te dimers. The gray/cyan spheres represent the undistorted/distorted positions of top Te atoms. (**f**) is the Fourier transform of the topographic image (including the 1/3 modulation) of IrTe$_2$ in the area shown in (**a**). A 1/6 peak near (010) is clearly seen. (**g**) The line profiles along the red and blue arrows show the peaks (1/6 and 1/3) associated with the $q=1/6$ modulation. The tunneling setup conditions for (**a**/**b**) are: $V_{bias} = 5$ mV and $I_{set} = 5/15$ nA.

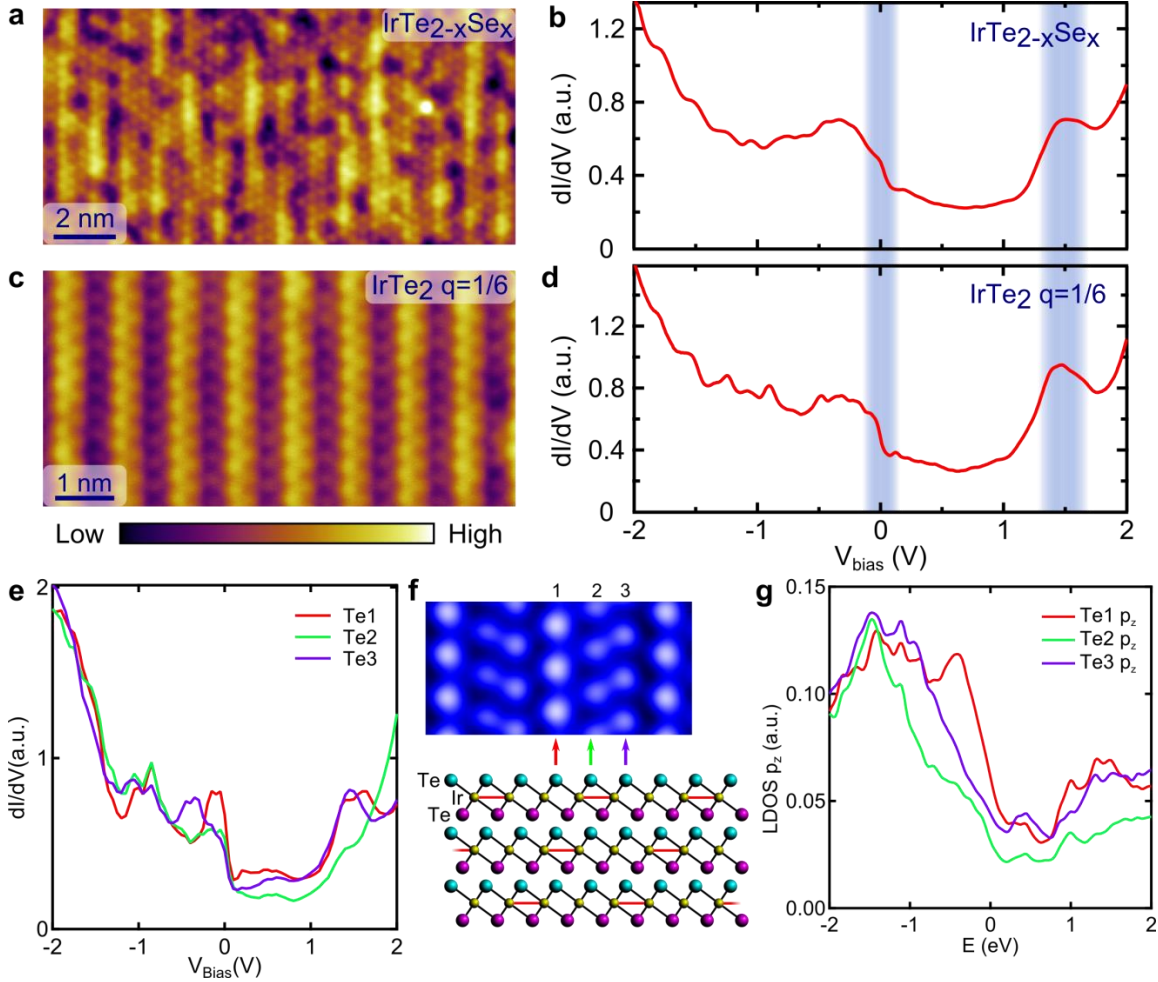

**FIG. 2** (Color online) **LDOS of IrTe$_{2-x}$Se$_x$ and IrTe$_2$ at 4.5 K. a** and **b** show typical topographic image (4.5 K) of IrTe$_{2-x}$Se$_x$ and IrTe$_2$, respectively. These suggest that the ground state of IrTe$_{2-x}$Se$_x$ is a long-range-ordered $q$=1/6 phase, even in the presence of Te/Se disorder. **c** and **d** show averaged $dI/dV$ spectroscopy data (LDOS) with large energy range of IrTe$_{2-x}$Se$_x$ and IrTe$_2$, respectively, indicating qualitatively identical LDOS. **e**, STS of individual atomic columns (defined in **f**) within half period of the 1/6 phase. **f**, unit-cell-averaged STM image with labels of atomic columns (1, 2, 3) and a cartoon of IrTe$_2$ lattice (side-view). The crystal orientation is determined by tracing the shift of the stripes across a step edge on the same surface (see Fig. **S4**). **g**, calculated LDOS of Te $p_z$ orbitals on the top Te layer (defined in **f**), while the $p_z$ orbital is pointing out of plane. Here the spectroscopic data were measured over a very wide high energy range (+/-2V).

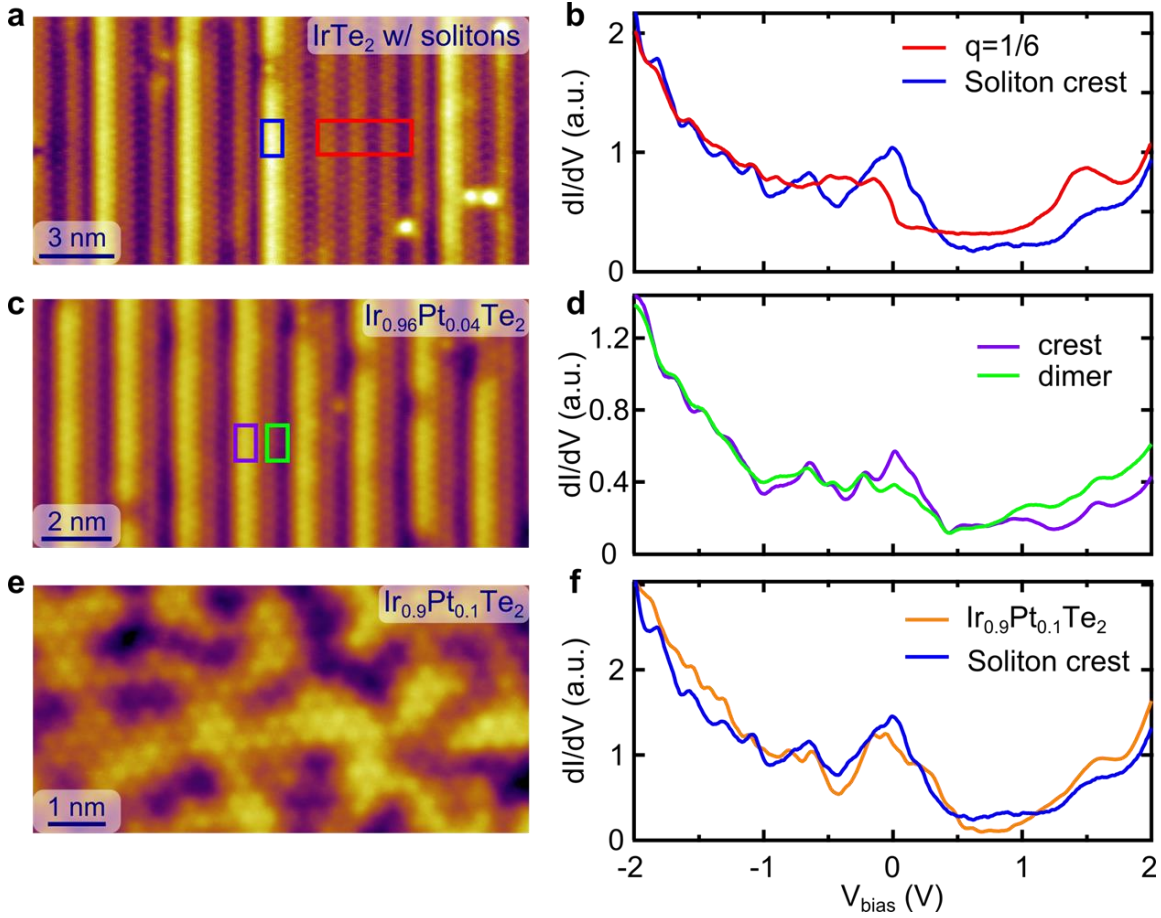

**FIG. 3** (Color online) **LDOS of soliton crest in IrTe$_2$ and Pt-doped IrTe$_2$ at 4.5 K. a**, diluted solitons occasionally observed in IrTe$_2$ at 4.5 K. **b**, areal averaged STS of soliton crest (blue) in comparison with that of 1/6 phase (red). **c**, typical topographic image of slightly Pt doped IrTe$_2$, where 1/5 phase is stabilized at 4.5 K. **d**, column-averaged STS data of crest (purple) and valley (green) in the 1/5 phase. The LDOS of soliton (crest) is at Fermi energy $E_F$ is higher than that of valley (dimerized stripes). **e**, typical topographic image of ~10% Pt doped IrTe$_2$. The stripe modulated phases is completely suppressed with significant Pt doping, resulting in the high temperature phase (1×1). **f**, averaged STS data of Ir$_{0.9}$Pt$_{0.1}$Te$_2$ (orange) in comparison with that of soliton crest in **a** (blue). The qualitative similarity between these data suggests that the soliton (crest) is undimerized atomic rows.

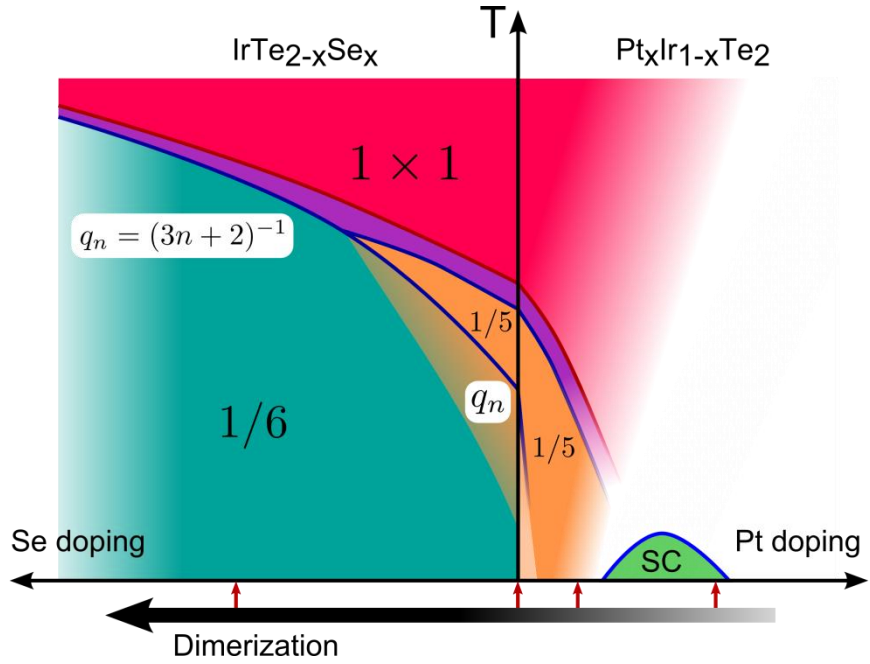

**FIG. 4** (Color online) **A unified-phase diagram of doped IrTe$_2$.** The first order phase transition $T_C$ is enhanced when Te is partially replaced by Se, suggesting that Se doping strengthens the Ir-Ir dimerization. The large hysteretic transition $T_S$ is suppressed by the enhancement of dimerization. In contrast, Pt doping weakens dimerization tendency so that $T_C$ is suppressed and superconductivity emerges at low temperature (~ 3 K).

# Hierarchical stripe phases in IrTe$_2$ driven by competition between Ir dimerization and Te bonding

Jixia Dai[1,2], Kristjan Haule[1], J.J. Yang[3], Y.S. Oh[1,2], S-W. Cheong[1,2,3] and Weida Wu[1,2]

[1] Department of Physics and Astronomy, Rutgers University, Piscataway, NJ, 08854, USA

[2] Rutgers Center for Emergent Materials, Rutgers University, Piscataway, NJ, 08854, USA

[3] Laboratory for Pohang Emergent Materials and Department of Physics,

Pohang University of Science and Technology, Pohang 790-784, Korea

**List of contents:**

**Part I. Removal of the stripe-related modulation from images taken at various bias voltages.**

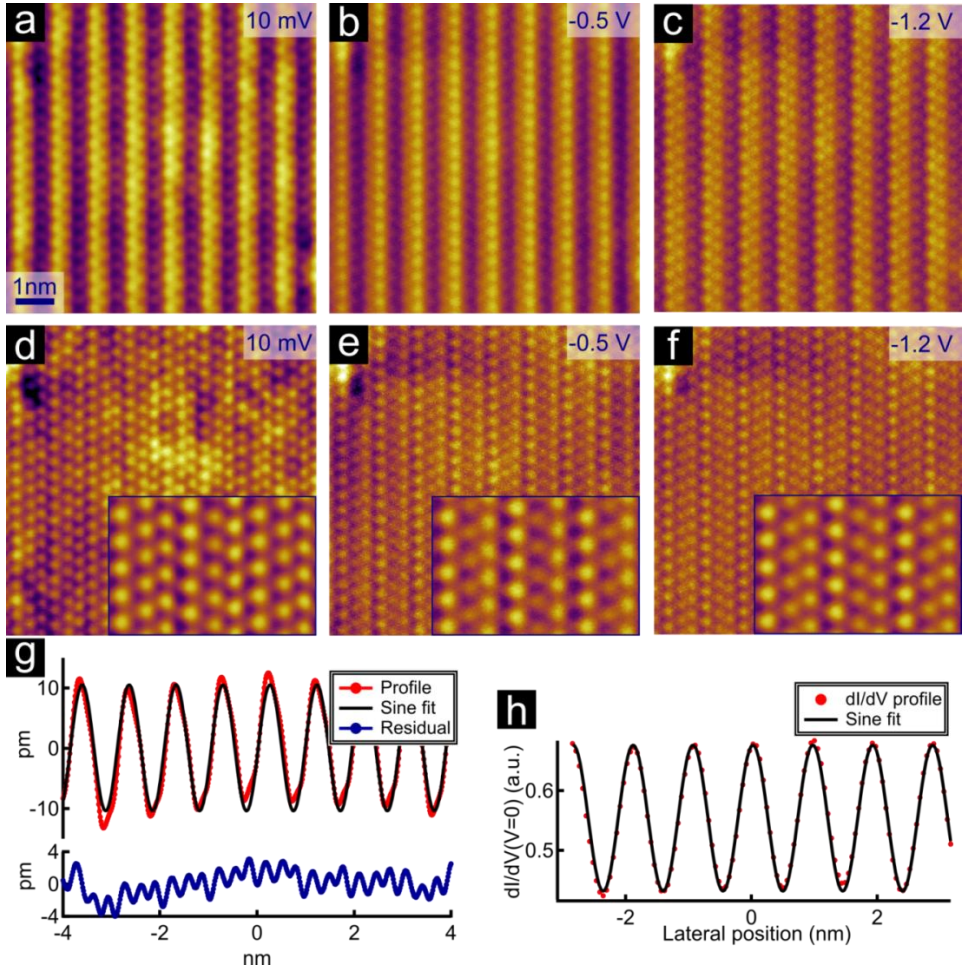

**Figure S1** Raw topographic images (**a-c**) taken over the exact same area that are largely affected by the stripe-related LDOS modulation and hence it is not possible to position the Te atoms using these images. By fitting the column-averaged profile to a sine function (upper panel in (**g**)) and subtracting this background from the raw images, we are able to obtain images (**d-f**) that are relatively consistent about the atomic positions. The reason for such a simple method to work is because the LDOS modulation are highly harmonic (**h**).

STM imaging are sensitive to both local atomic positions and local density of states (LDOS), in IrTe$_2$ the raw images (supplementary Fig. **S1a-c**) are dominated by contributions from the electronic modulation, which makes the positioning of the atoms difficult. However, the electronic modulation is highly harmonic (Fig. **S1h**), and by removing a simple sinusoidal component from these images we can effectively minimize the contribution from the DOS modulation and thus position the Te atoms accurately (Fig **1a-b**). This method has been tested with images taken with very different bias voltages, while yielding essentially similar results (Fig. **S1e-f**). Furthermore, by averaging the images with the 1x6 unit cells, we can remove the inhomogeneity related to local dopants and enhance the visibility of the Te-Te dimers (insets in Fig. **S1e-f**).

**Part II. Typical image of large area *q*=1/6 modulation at 4.5 K.**

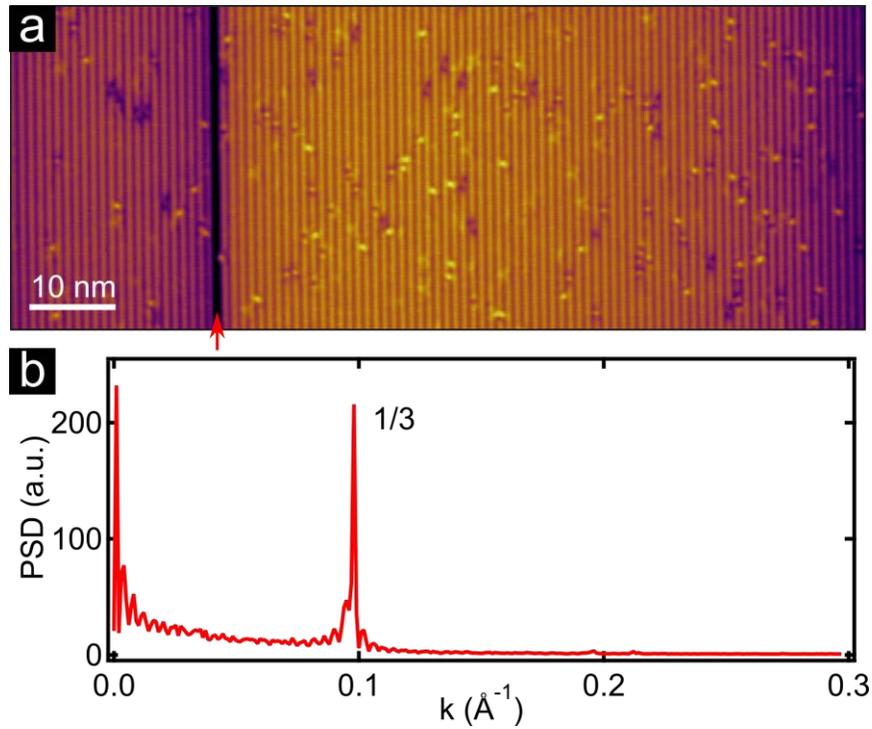

**Figure S2 (a)** Large area (100 nm) with *q*=1/6 modulation except for one dark anti-phase boundary indicated by the red arrow. Each vertical stripe corresponds to three atomic rows as shown in Fig 1 in the main text. **(b)** One-dimensional Fourier transform of **(a)** showing a very strong first order harmonic peak and an almost invisible second order peak.

Shown in Fig. **S2** is a large image of *q*=1/6 modulation taken at $T = 4.5K$. It is the dominant phase at low temperature. This image was taken with $V_{bias} = -200$ mV and $I_{set} = 100$ pA. It does not have atomic resolution and every stripe in the image is about 1 nm wide, corresponding to three atomic columns as shown in Fig. **1** of main text.

**Part III. Image of *q*=1/5 modulation taken at 77K and the corresponding modulation removal.**

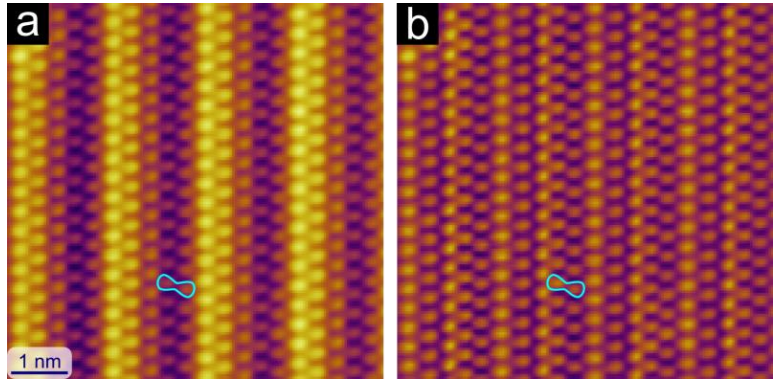

**Figure S3** Topographic image of the *q*=1/5 phase measured at *T*=77K. (**a**) is the raw image taken with $V_{bias}$ = 10 mV and $I_{set}$ = 2 nA. (**b**) is the same image but with the first harmonic of the modulation removed. The cyan colored curves indicate the location of the Te-Te dimer.

We did not observe large areas of quenched *q*=1/5 phase at *T* = 4.5 K. However, large patches of 1/5 phase do exist at *T* = 77 K. Shown in Fig. **S3a** is an area of *q*=1/5 phase taken at 77 K, while in Fig. **S3b** we show the same image after removing a sinusoidal background. The STM image of the *q*=1/5 modulation is rather aharmonic and therefore this simple modulation removal left some more residual than the 1/6 case. Nonetheless, the Te dimers are still visible after the modulation removal.

**Part IV. Image across an atomic step edge that allows for determination of crystallographic orientation.**

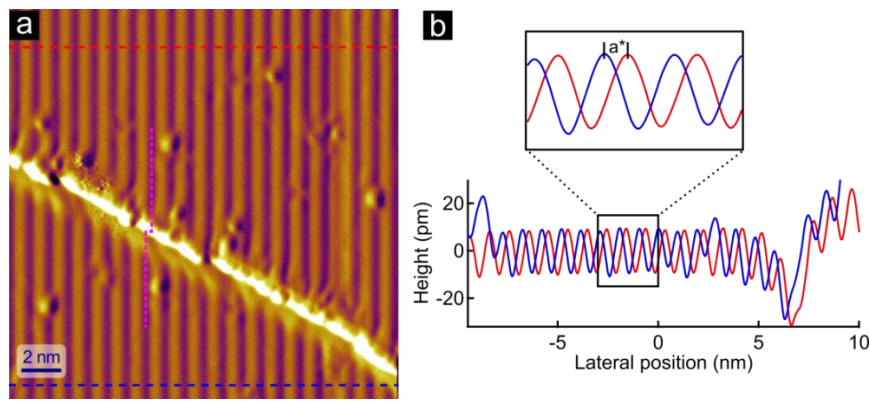

**Figure S4 (a)** Topographic image (derivative) of the $q=1/6$ modulation across a step edge shows the shift of one atomic column between the two adjacent layers, indicated by the pink dashed line. The step edge is about 5.4 Å high, corresponding to one Te-Ir-Te trilayer. **(b)** Profiles along the red and blue dashed lines in **(a)** showing the shift of one atomic column between them. The red curve is shifted down by 5.4 Å to align with the blue curve.

Atomic step edges of 5.4 Å could be found on the surface of cleaved $IrTe_2$. These step edges correspond to single trilayers of Te-Ir-Te. As mentioned earlier, the stripe modulations cut across the layers in a diagonal direction. Only knowing about the surface layer, one cannot decide whether the modulation is going to the left or right. In a word, $IrTe_2$ crystal is three-fold symmetric, while the surface Te layer is six-fold symmetric. When the stripes form, the crystal loses all of its rotation symmetry but the surface still appears to be two-fold symmetric. With these step edges, one can easily determine the orientation of the crystal. In the case shown in Fig. **S4**, we can see that all of the features shift to the left when one traces from the top layer (upper half of Fig. **S4a**) to the next deeper layer (lower half Fig. **S4a**). This crystallographic convention is used for all data shown in this study.

**Part V. Typical image and spectroscopy of Au(111) surface in the tip calibration.**

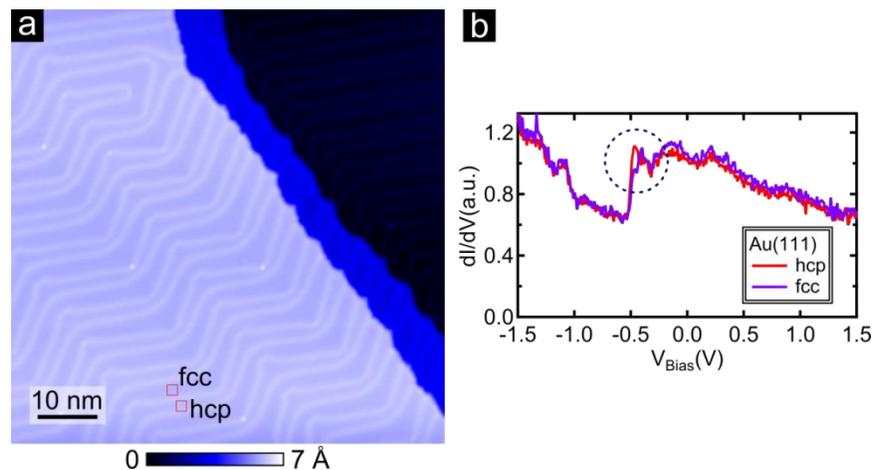

**Figure S5** | Tip calibration. (a) Au(111) surface showing nice herringbone reconstruction with two monoatomic step edges of ~2.4 Å each. (b) *dI/dV* measurements on Au(111) show very clear steps near -0.5 V corresponding to the emergence of the Shockley surface states. The difference between the fcc and hcp sites could be seen near -0.5 V.

Electrochemically etched tungsten tips are used in our experiment. The tips are firstly treated with field emission and then refined by voltage pulsing while scanning on Au(111) surface. The tips are treated until the following criteria are satisfied. First, the tip could image the herring bone structure with sharp contrast as shown in Fig. **S5a**. Second, the *dI/dV* measured show the step-like Shockley surface states near -0.5 eV together with the small but observable difference between the fcc and hcp sites (dashed circle), following Chen *et al.* (Ref. 1).